\documentclass[twocolumn,showpacs,preprintnumbers,amsmath,amssymb]{revtex4}
\usepackage{graphicx}
\usepackage{dcolumn}
\usepackage{bm}

\begin{document}

\preprint{APS/123-QED}

\title{Dwell times for transmission and reflection}

\author{N. L. Chuprikov}
\email{chnl@tspu.edu.ru}
\affiliation{%
Tomsk State Pedagogical University, 634041, Tomsk, Russia}
\altaffiliation[Also at
]{Physics Department, Tomsk State University.}

\date{\today}

\begin{abstract}

As was shown in quant-ph/0405028, the state of a tunneling particle can be uniquely
presented as a coherent superposition of two states to describe alternative
sub-processes, transmission and reflection. In this paper, on the basis of the
stationary wave functions for these sub-processes, we give new definitions of the
dwell times for transmission and reflection. In the case of rectangular potential
barriers the dwell times are obtained explicitly. In contrast with the well-known
B\"{u}ttiker's dwell-time, our dwell time for transmission increases exponentially,
for the under-barrier tunneling, with increasing the barrier's width. By our
approach the well-known Hartman effect is rather an artifact resulted from an
improper interpretation of the wave-packet tunneling and experimental data, but not
a real physical effect accompanying the tunneling phenomenon.

\end{abstract}
\maketitle
\newcommand {\uta} {\tau_{tr}}
\newcommand {\utb} {\tau_{ref}}
\newcommand{\ppp}{\mbox{\hspace{5mm}}}
\newcommand{\ooo}{\mbox{\hspace{3mm}}}
\newcommand{\ooa}{\mbox{\hspace{1mm}}}

\section{Introduction}

In this paper we continue our study \cite{Ch5} of the temporal aspects of tunneling
considered in our approach as a combined stochastic process to consist from two
alternative elementary ones, transmission and reflection, evolved coherently. In
\cite{Ch5} we have introduced the transmission and reflection times provided that a
particle is in a localized state. These time scales were extracted from the time
dependence of the average positions of transmitted and reflected particles.

It is evident that this way of timing a particle is not applicable in the limiting
case, when the particle is in some stationary state described by a non-normalized
wave function. This case is of great importance, since it is connected to the
scattering of uniform beams of identically prepared particles, noninteracting with
each other. To find a time spent by transmitted and reflected particles in the
barrier region, in the stationary scattering problem, is the main aim of the present
paper.

\section{Stationary wave functions for transmission and reflection in the case of
rectangular potential barriers}

How to obtain the wave functions for transmission and reflection, for a given
potential and initial wave function, is presented in \cite{Ch5}. Here we expose both
the functions in the form to be more suitable for our purpose.

Let setting the scattering problem be such as in \cite{Ch5}. In particular, we will
suppose that the rectangular potential barrier of height $V_0$ is localized in the
interval $[a,b]$. Then the wave functions for transmission, $\Psi_{tr}(x;E)$, and
reflection, $\Psi_{ref}(x;E)$, can be written for $E\le V_0$ as follows ($E$ is the
energy of a particle).

For $x\le a$
\begin{eqnarray} \label{1}
\Psi_{tr}=A_{in}^{tr} e^{ikx};\ooo
\Psi_{ref}=A_{in}^{ref}e^{ikx}+b_{out}e^{-ik(x-2a)};
\end{eqnarray}
for $a\le x\le x_c$
\begin{eqnarray} \label{2}
\Psi_{tr}=a^l_{tr}\sinh(\beta)+b^l_{tr}\cosh(\beta);\nonumber\\
\Psi_{ref}=a^l_{ref}\sinh(\beta);
\end{eqnarray}
for $x_c\le x\le b$
\begin{eqnarray} \label{3}
\Psi_{tr}=a^r_{tr}\sinh(\beta)+b^r_{tr}\cosh(\beta);\ooo \Psi_{ref}=0;
\end{eqnarray}
for $x\ge b$
\begin{eqnarray} \label{4}
\Psi_{tr}=a_{uot}e^{ik(x-d)};\ooo \Psi_{ref}=0;
\end{eqnarray}
where $d=b-a$, $x_c=(b+a)/2$, $\beta=\kappa(x-x_c)$,$k=\sqrt{2mE/\hbar^2},$
$\kappa=\sqrt{2m(V_0-E)/\hbar^2};$ $m$ is the mass of a particle. Besides,
\[A_{in}^{tr}=a^*_{out}\left(a_{out}+b_{out}\right),\]
\[A_{in}^{ref}=b_{out}\left(b^*_{out}-a^*_{out}\right);\]
\[a^l_{tr}=\frac{i}{\kappa}PA_{in}^{tr}e^{ika},\ooo
b^l_{tr}=\frac{1}{\kappa}QA_{in}^{tr}e^{ika},\]
\[a^r_{tr}=\frac{i}{\kappa}P^*a_{out}e^{ika},\ooo
b^r_{tr}=\frac{1}{\kappa}Q^*a_{out}e^{ika},\]
\[a^l_{ref}=\frac{i}{\kappa}\left(PA_{in}^{ref}-P^*b_{out}\right)e^{ika};\]
\[a_{out}=\frac{1}{2}\left(\frac{P}{P^*}+\frac{Q}{Q^*}\right);\ooo
b_{out}=\frac{1}{2}\left(\frac{P}{P^*}-\frac{Q}{Q^*}\right).\]  In its turn,
\[P=k\cosh(\varphi)-i\kappa \sinh(\varphi),\] \[Q=\kappa
\cosh(\varphi)+ik\sinh(\varphi);\ooo \varphi=\frac{\kappa d}{2}.\]

One can easily check that
\[\frac{a_{out}}{A_{in}^{tr}}=-\frac{b_{out}}{A_{in}^{ref}}=\frac{Q}{Q^*};\]
and $|a_{out}|^2=T$, $|b_{out}|^2=R$ where $T$ and $R$ are the coefficients of
transmission and reflection, respectively ($T+R=1$); $A_{in}^{tr}+A_{in}^{ref}=1$.
Thus, $b^l_{tr}=b^r_{tr}=b_{tr}$.

\section{Dwell time for transmission}

Note, both the stationary wave functions are non-normalized and, hence, the standard
timing procedure becomes usefulness in timing a particle with the well defined
energy. At the same time the properties of these wave functions provide another
natural way for timing a particle in the barrier region.

Indeed, the density of the probability flux, $I_{tr}$, for $\Psi_{tr}(x;E)$ is
everywhere constant and equals to $\frac{\hbar k}{m}T$. This means that the velocity
of this flux, $v_{tr}$, can be written as
\[v_{tr}(x)=\frac{I_{tr}}{|\Psi_{tr}(x;E)|^2}.\]
Taking into account Exps. (\ref{1}) - (\ref{4}) for $\Psi_{tr}(x;E)$, we see that
outside the barrier region $v_{tr}=\hbar k/m$. In the barrier region, the velocity of
the flux decreases exponentially when it approaches the midpoint $x_c$. One can easily
show that $|\Psi_{tr}(a;E)|=|\Psi_{tr}(x;E)|=\sqrt{T}$, but
$|\Psi_{tr}(x_c;E)|=\sqrt{T}|Q|/\kappa$.

Thus, any selected infinitesimal element of the flux passes the barrier region for
the time $\tau^{tr}_{dwell}$, where
\begin{eqnarray} \label{5}
\tau^{tr}_{dwell}=\frac{1}{I_{tr}}\int_a^b|\Psi_{tr}(x;E)|^2 dx.
\end{eqnarray}

One can easily show that
\begin{eqnarray} \label{6}
\int_a^b|\Psi_{tr}(x;E)|^2 dx=\frac{d}{4}\left(2|b_{tr}|^2-|a^r_{tr}|^2-
|a^l_{tr}|^2\right)\nonumber\\ +\frac{\sinh(2\varphi)}{4\kappa} \left(2|b_{tr}|^2
+|a^r_{tr}|^2+ |a^l_{tr}|^2\right)\nonumber\\ +\frac{\sinh^2(\varphi)}{\kappa}
Re[(a^r_{tr}-a^l_{tr})b^*_{tr}].
\end{eqnarray}
Then, taking into account Exps. (\ref{1}) - (\ref{4})), we obtain
\[2|b_{tr}|^2-|a^r_{tr}|^2-|a^l_{tr}|^2=\frac{2}{\kappa^2}(\kappa^2-k^2)T,\]
\[2|b_{tr}|^2+|a^r_{tr}|^2+|a^l_{tr}|^2= \frac{2}{\kappa^2}(\kappa^2+k^2)
T\cosh(2\varphi),\]
\[Re[(a^r_{tr}-a^l_{tr})b^*_{tr}]=-\frac{1}{\kappa^2}(\kappa^2+k^2)T\sinh(2\varphi).\]
Eventually, we have
\begin{eqnarray} \label{7}
\tau^{tr}_{dwell}=\frac{m}{2\hbar k\kappa^3}\left[\left(\kappa^2-k^2\right)\kappa d
+\kappa_0^2 \sinh(\kappa d)\right];
\end{eqnarray}
where
$\kappa_0=\sqrt{2m|V_0|/\hbar^2}.$

Similarly, for $E\ge V_0$
\begin{eqnarray} \label{9}
\tau^{tr}_{dwell}=\frac{m}{2\hbar k\kappa^3}\left[\left(\kappa^2+k^2\right)\kappa d
-\theta \kappa_0^2 \sin(\kappa d)\right];
\end{eqnarray}
where $\kappa=\sqrt{2m(E-V_0)/\hbar^2};$ $\theta=1$, if $V_0>0$; otherwise,
$\theta=-1.$

\section{Dwell time for reflection}

As is seen, the above time scale for transmission (\ref{5}) (though the way of
introducing is different) coincides by form with the well-known B\"{u}ttiker's dwell
time (see \cite{But}). However, unlike the latter, Exp. (\ref{5}) describes only
transmitted particles.

Doing in the spirit of the B\"{u}ttiker's dwell-time concept, we introduce the dwell
time for reflection, $\tau^{ref}_{dwell}$, as
\begin{eqnarray} \label{14}
\tau^{ref}_{dwell}=\frac{1}{I_{ref}} \int_a^{x_c}|\Psi_{ref}(x)|^2 dx;
\end{eqnarray}
where $I_{ref}=\frac{\hbar k}{m}R$ is the incident probability flux (note, the
probability flux for $\Psi_{ref}(x)$ itself is zero).

Considering the expressions for $\Psi_{ref}(x)$ one can easily show that for $E<V_0$
\begin{eqnarray} \label{30}
\tau^{ref}_{dwell}=\frac{m k}{\hbar \kappa}\cdot\frac{\sinh(\kappa d)-\kappa
d}{\kappa^2+\kappa^2_0 \sinh^2(\kappa d/2)};
\end{eqnarray}
for $E\ge V_0$
\begin{eqnarray} \label{31}
\tau^{ref}_{dwell}=\frac{m k}{\hbar \kappa}\cdot\frac{\kappa d-\sin(\kappa
d)}{\kappa^2+\theta\kappa^2_0 \sin^2(\kappa d/2)}.
\end{eqnarray}

\section{Our and B\"{u}ttiker's concepts of the dwell time: Comparative analysis}

As was shown in \cite{Ch2}, the phase-time and old dwell-time concepts,
$\tau_{dwell}$, do not imply the passage to the case of a free particle, that is,
when $\kappa_0 d\to 0$. At the same time, our concept of the asymptotic times for
transmission and reflection (see \cite{Ch5}) ensures this passage. One can easily
check that the above dwell time for transmission implies this passage, too. Note,
both the characteristic times are nonnegative quantities.

Let $\tau_{free}(k)$ be the time spent by a free particle in the interval $[a,b]$:
$\tau_{free}(k)=\frac{md}{\hbar k}$. Then one can show that for barriers, in the limit
$k\to 0$ ($\kappa_0 d\neq 0$), we have
\begin{eqnarray} \label{15}
\frac{\tau^{tr}_{dwell}}{\tau_{free}}=\frac{1}{2}\left[1+\frac{\sinh(\kappa_0
d)}{\kappa_0 d}\right];
\end{eqnarray}
for wells,
\begin{eqnarray} \label{16}
\frac{\tau^{tr}_{dwell}}{\tau_{free}}=\frac{1}{2}\left[1+\frac{\sin(\kappa_0
d)}{\kappa_0 d}\right].
\end{eqnarray}
Thus, not only for $k\neq 0$, but also for $k=0,$ $\tau^{tr}_{dwell}\to \tau_{free}$
when $\kappa_0 d\to 0.$ Remind (see \cite{Ch2}), the $\tau_{dwell}/ \tau_{free}=0$ in
the latter case.

Remind, the B\"{u}ttiker's dwell time for $E<V_0$ read as (see \cite{But})
\begin{eqnarray} \label{20}
\tau_{dwell}=\frac{m k}{\hbar\kappa}\cdot \frac{2\kappa d
(\kappa^2-k^2)+\kappa_0^2\sinh(2\kappa d)} {4k^2\kappa^2+ \kappa_0^4\sinh^2(\kappa
d)};
\end{eqnarray}
for $E\ge V_0$
\begin{eqnarray} \label{21}
\tau_{dwell}=\frac{m k}{\hbar\kappa}\cdot \frac{2\kappa d
(\kappa^2+k^2)-\kappa_0^2\sin(2\kappa d)} {4k^2\kappa^2+ \kappa_0^4\sin^2(\kappa d)}.
\end{eqnarray}

It is useful to compare formally a new dwell-time scale (to describe transmitted and
reflected particles separately) with B\"{u}ttiker's dwell time to describe all
particles jointly.

Figs.1 and 2 show the $E$-dependence of $\tau^{tr}_{dwell}/\tau_{free}$ and
$\tau_{dwell}/\tau_{free}$ for $V_0=-0.1eV$, $d=30nm$ and $V_0=0.1eV$, $d=15nm$,
respectively. Figs. 3 and 4 show the dependence of $\tau_{dwell}$,
$\tau^{tr}_{dwell}$ and $\tau_{free}$ on the barrier's width $d$ for $V_0=0.1eV$,
$E=0.11eV$ and $V_0=0.1eV$, $E=0.09eV$, respectively. In all cases, $m=0.067m_e$
where $m_e$ is the mass of an electron.

As is seen, $\tau^{tr}_{dwell}$ and $\tau_{dwell}$ coincide at the points of
resonance. However, this result is expected. Besides, $\tau^{tr}_{dwell}(E,d)$ is
more smooth function than $\tau_{dwell}(E,d).$ Moreover, except for some
neighborhoods of resonant points, $\tau^{tr}_{dwell}>\tau_{dwell}.$

Our results show that the most essential difference between these time scales takes
place for $E<V_0$ (see Figs.2 and 4). In particular, Fig.4 shows that
$\tau^{tr}_{dwell}$ increases exponentially when the width of an opaque potential
barrier increases. At the same time $\tau_{dwell}(E,d)$ (as well as the well-known
phase tunneling time and our asymptotic times for transmission and reflection)
tends, in this case, to some constant value.

As regards $\tau^{ref}_{dwell}$ (see (\ref{30}), (\ref{31})), in this limit
\[\tau^{ref}_{dwell}=\frac{2m k}{\hbar \kappa \kappa_0^2}.\] Note also,
$\tau^{ref}_{dwell}\to 0,$ for wells ($V_0<0$), when $k\to 0.$ However, in the case
of resonance, when $\kappa_0 d=\pi+2n\pi,$ $\tau^{ref}_{dwell}\to \infty$ at $k\to
0.$

\section{Scenario of tunneling a particle through an opaque rectangular barrier}

As is known, the just mentioned property of the phase and dwell tomes is excepted to
lead to superluminal velocities of a particle in the barrier region (the well-known
Hartman effect). However, we see that the behavior of $\tau^{tr}_{dwell}$, in
contrast to the B\"{u}ttiker's dwell time, denies the existence of the Hartman
effect for a particle with the well-defined energy. However, from our study it
follows that this effect does not appear also for a particle whose initial state is
described by the wave packet of any width. Fig.5 shows the expectation value of the
particle's position as a function of $t$. It was calculated for the transmitted wave
packet providing that $a=200nm$, $b=215nm$, $V_0=0.2 eV$. At $t=0$ the (combined)
state of the particle is described by the Gaussian wave packet peaked around $x=0$;
its half-width $10nm$; the average energy of the particle $0.05eV$.

This figure displays explicitly the difference between the exact and asymptotic
transmission times. While the former gives just the (average) time spent by a
particle in the barrier region, the latter describes, in fact, the (average) lag or
outstripping of a transmitted particle with respect to a free particle whose average
velocity equals to the asymptotic average velocity of a scattered particle.

In the case considered (see Fig.5) the exact transmission time equals approximately
to $0.155ps$, the asymptotic one is of $0.01ps$, and $\tau_{free}\approx 0.025ps$.
As is seen, the dwell time for transmission and exact transmission time, both
evidence that, though the asymptotic transmission time is small for this opaque
barrier, transmitted particles spend much time in the barrier region.

Note, the possibility of accelerating a particle in the region of an opaque potential
barrier, to superluminal velocities, is doubtful {\it a priori}. The point is that the
average energy of a particle, $\overline{\hat{H}}_{tr}$, is constant for this static
potential barrier. Thus, for the average kinetic ($\overline{\hat{K}}_{tr}$ ) and
potential ($\overline{\hat{V}}_{tr}$) energies of a particle we have, for any $t$,
\[\overline{\hat{K}}_{tr}+ \overline{\hat{V}}_{tr} =\overline{\hat{K}}_{0,tr},\]
where $\overline{\hat{K}}_{0,tr}$ is the average kinetic energy of transmitted
particles at $t=0$.

It is evident that $\overline{\hat{V}}_{tr}>0,$ if $V_0>0$. Hence, at the stage of
scattering, $\overline{\hat{K}}_{tr}$ may only decrease. Of course, this fact does not
at all mean that the average velocity of a particle, at the stage of scattering,
cannot be larger than its asymptotic value. The point is that
\[\overline{\hat{K}}_{tr}=\frac{\overline{p}_{tr}^2}{2m}+
\frac{\overline{(\delta p)^2}}{2m};\] $p$ is the particle's momentum. Thus,
$\bar{p}_{tr}$ may increase due to diminishing the wave-packet's width in the
momentum space. Our calculations show that this takes place near the barrier region
(see Fig.5).

Thus, when approaching the left boundary of the barrier, the particles is accelerated,
on the average. As a result, it may overcome the potential barrier. When the particle
enters the barrier region, it almost stops. The region of an opaque potential barrier
serves as a storage of particles. Behind the barrier the average velocity of
particles may first exceed its own asymptotic value. However, far from the barrier,
when the width of the wave packet, in the momentum space, backs to its initial
(asymptotic) value, the same does the average velocity of particles.

Note, the more is the part of $\frac{\overline{(\delta p)^2}}{2m}$ in the average
kinetic energy of a particle at $t=0$, the larger may be the grows of the particle's
velocity near the barrier region. However, one has to bear in mind that this velocity
cannot essentially exceed the value, $\sqrt{2mV_0}/m$. So that, as it follows from our
approach, tunneling a particle through an opaque potential barrier must not be
accompanied by the Hartman effect.

\section{Conclusion}

So, in our approach, the state of the whole quantum ensemble of tunneling particles
is considered as a coherent superposition of the states of the subensembles of
transmitted and reflected particles. For a given potential and initial state of a
particle, there is an unique pair of solutions to the one-dimensional Schr\"odinger
equation, which describe both the subensembles at all stages of tunneling. For each
of these subensembles we have introduced three time scales. Two of them (the exact
and asymptotic tunneling times) describe a completed scattering process. The third
time scale (the dwell time) relates to the stationary tunneling problem. It should
be stressed that the asymptotic (transmission and reflection) times differ from the
well-known phase asymptotic times. The dwell times for transmission and reflection
differ from the B\"{u}ttiker's dwell time. Analysis of tunneling a particle through
an opaque rectangular barrier, carried out in the framework of our approach, does
not confirm the existence of the Hartman effect.

At the end we have to point to some shortcomings of the above time concepts. So, the
concept of the exact tunneling times does not give explicit expressions for these
quantities; besides, it is not suitable for timing a particle with the well-defined
energy (this becomes more evident in the case of the reflection of wide (in
$x$-space) wave packets, when the expectation value of the particle's position
simply does not "enter" the barrier region). The main shortcoming of the dwell-time
concept is that it is valid only for the stationary scattering process.

The our next purpose is to present the concept of the Larmor times for transmission
and reflection. By our opinion, this concept is of great importance in solving the
tunneling time problem.

\section*{Figure captions}

Fig. 1. $\tau_{dwell}/\tau_{free}$ and $\tau^{tr}_{dwell}/\tau_{free}$ versus $E$,
for $d=30nm$ and $V_0=-0.1 eV$.

\vspace{1cm} \noindent Fig. 2. $\tau_{dwell}/\tau_{free}$ and
$\tau^{tr}_{dwell}/\tau_{free}$ versus $E$, for $d=15nm$ and $V_0=0.1 eV$.

\vspace{1cm} \noindent Fig. 3. $\tau_{dwell},$ $\tau^{tr}_{dwell}$ and $\tau_{free}$
versus $d$, for $E=0.11eV$ and $V_0=0.1 eV$.

\vspace{1cm} \noindent Fig. 4.  $\tau_{dwell},$ $\tau^{tr}_{dwell}$ and
$/\tau_{free}$ versus $d$, for $E=0.09eV$ and $V_0=0.1 eV$.

\vspace{1cm} \noindent Fig. 5. The $t$-dependence of the average position of
transmitted particles; the initial (full) state vector represents the Gaussian wave
packet peaked around the point $x=0$, its half-width equals to $10nm,$ the average
kinetic particle's energy is $0.05eV;$ $a=200nm$, $b=215nm$.
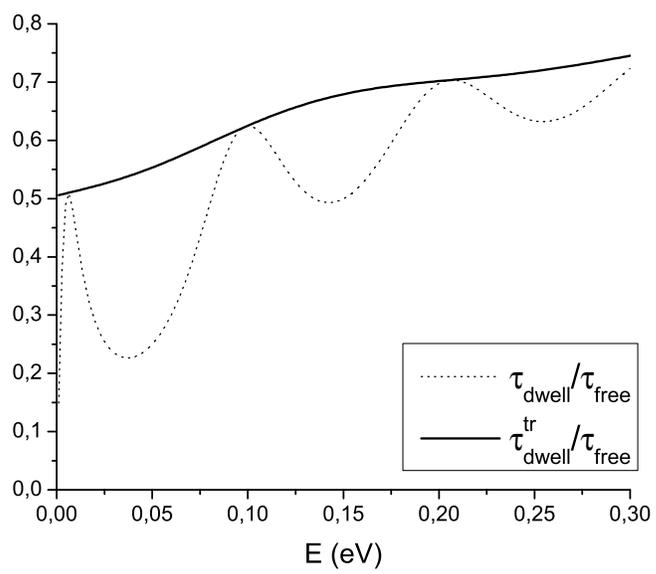
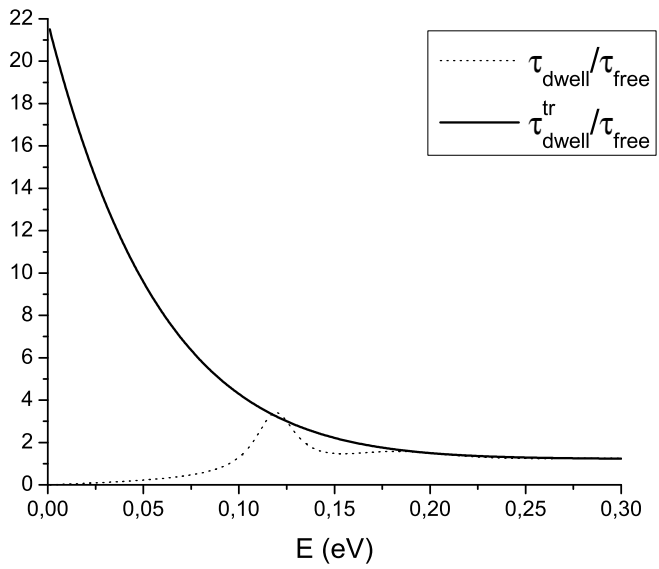
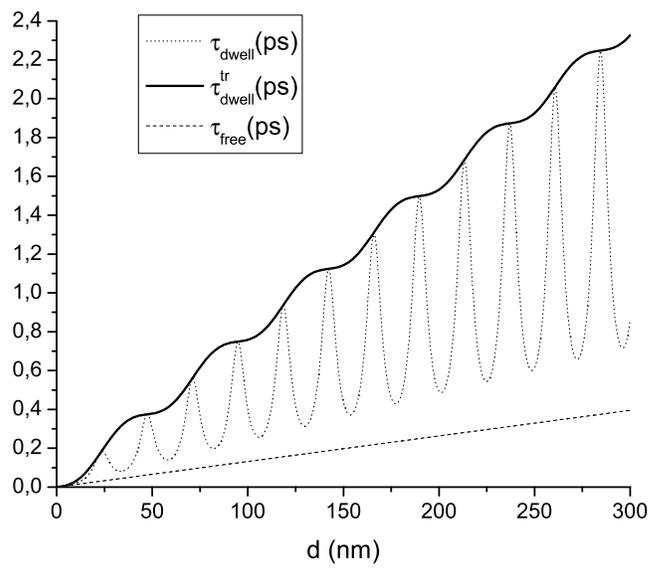
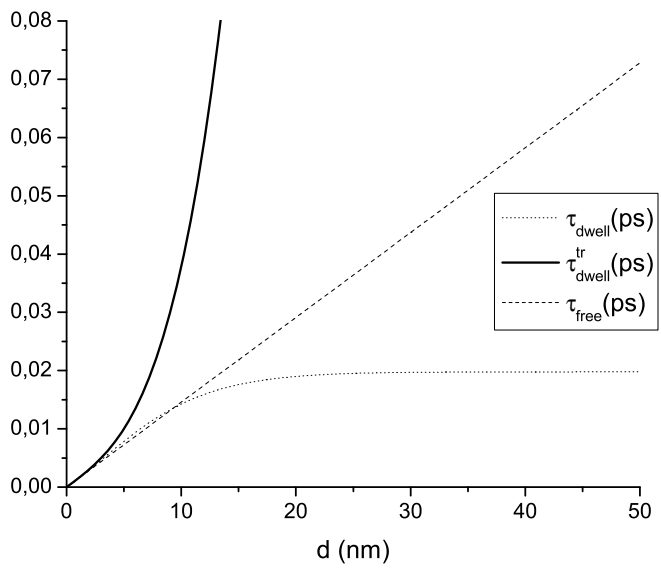
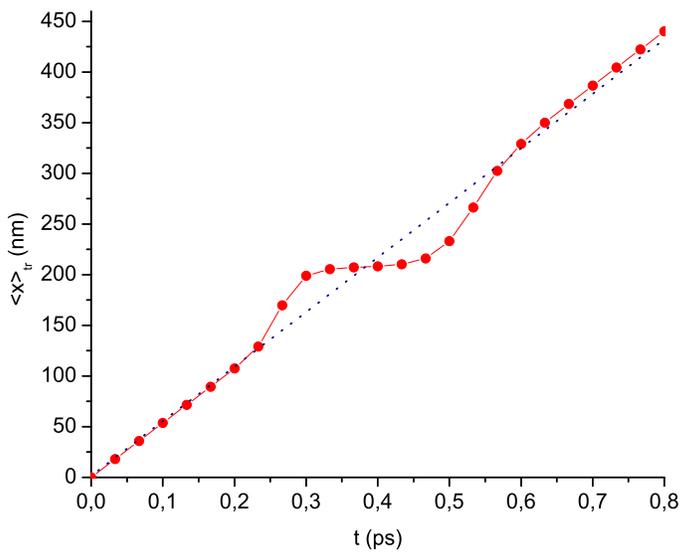
\end{document}